%% file: ms_axion_search_Lett.tex
\documentclass{ptephy_v2}%%%%where ptephy_v1 is the template name
\usepackage{hyperref}
\usepackage{aas_macros}
\newcommand{\bfB}{\boldsymbol{B}}

\newcommand{\bfJ}{\boldsymbol{J}}
\newcommand{\bfPhi}{\boldsymbol{\Phi}}
\newcommand{\bfr}{\boldsymbol{r}}
\newcommand{\gag}{g_{\rm a\gamma}}
\newcommand{\mass}{m_{\rm a}}
\newcommand{\unit}[1]{\,\rm{#1}}

\begin{document}

\title{Hunting axion dark matter signatures in low-frequency terrestrial magnetic fields}

%%%% To generate auto affiliation numbers please use \author{}\affil{} command

\author[1,2,3]{Atsushi Taruya}
\affil[1]{Center for Gravitational Physics and Quantum Information, Yukawa Institute for Theoretical Physics, Kyoto University, Kyoto 606-8502, Japan \email{ataruya@yukawa.kyoto-u.ac.jp}}

\affil[2]{Kavli Institute for the Physics and Mathematics of the Universe, Todai Institutes for Advanced Study, The University of Tokyo, (Kavli IPMU, WPI), Kashiwa, Chiba 277-8583, Japan}

\affil[3]{Korea Institute for Advanced Study, 85 Hoegiro, Dongdaemun-gu, Seoul 02455, Republic of Korea}

\author[4,5,6]{Atsushi Nishizawa}

\affil[4]{Physics Program, Graduate School of Advanced Science and Engineering, Hiroshima University, Higashi-Hiroshima, Hiroshima 739-8526, Japan}

\affil[5]{Astrophysical Science Center, Hiroshima University, Higashi-Hiroshima, Hiroshima 739-8526, Japan}

\affil[6]{Research Center for the Early Universe (RESCEU), Graduate School of Science, The University of Tokyo, Tokyo 113-0033, Japan}

\author[7]{Yoshiaki Himemoto}

\affil[7]{Department of Liberal Arts and Basic Sciences, College of Industrial Technology, Nihon University, Narashino, Chiba 275-8576, Japan}

%%% To include the collaborator name... Please use the command "\collaborator"
%%% For example: \collaborator{ATLAS Collaboration}

\begin{abstract}%
We show that Earth's natural environment can serve as a powerful probe for ultralight axion dark matter. In the presence of global geomagnetic fields, the axions with masses ranging from $10^{-15}$\,eV--$10^{-13}$\,eV induce electromagnetic waves in the (sub-) extremely low-frequency band ($0.3-30$\,Hz) through the axion-photon coupling. We predict the amplitude of induced magnetic fields in the Earth-ionosphere cavity, taking the finite conductivity of the atmosphere into account. This allows us to constrain the axion-photon coupling parameter, $\gag$, from the long-term monitoring data of the low-frequency magnetic fields, resulting in a significant improvement from the previous constraints down to $\gag \lesssim 4\times10^{-13}\,{\rm GeV}^{-1}$ for axion mass $\sim 3 \times 10^{-14}\unit{eV}$.
\end{abstract}

\subjectindex{xxxx, xxx}

\maketitle

%%%%%%%%%%%%%%%%%%%%%%%%%%%%%%%%%%%%%%%%%%%%%%%%%%%%%%%%
\noindent{\bf Introduction.} Many independent astronomical observations indicate that the matter content of our Universe is dominated by an invisible matter component, referred to as dark matter (DM). DM cannot be described by the standard model of particle physics, and its microscopic origin is one of the greatest mysteries in cosmology and fundamental physics. Among the various DM candidates, axion or axion-like particles, initially introduced to address the strong CP problem in QCD~\cite{peccei_quinn,Weinberg:1977ma,Wilczek:1977pj}, have long been a focal point of interest as representatives of the ultralight DM candidates, with masses of the axion DM $\mass$ ranging from $10^{-23}{\rm eV}$ to $1\,{\rm eV}$ (e.g., \cite{Preskill:1982cy,Abbott:1982af,Dine:1982ah,Svrcek_Witten2006,Arvanitaki_etal2010,Hui_etal2017}, see also \cite{Marsh_review2016} for a review). 

The axion DM can couple with electromagnetic (EM) fields, described by the interaction Lagrangian, $\mathcal{L}_{\rm int}=(\gag/4)\,a\,F_{\mu\nu}\tilde{F}^{\mu\nu}$, where $a$ is the axion field, $F_{\mu\nu}$ is the EM field strength tensor with its dual given by $\tilde{F}^{\mu\nu}=\epsilon^{\mu\nu\alpha\beta}F_{\alpha\beta}/2$, and $\gag$ represents the coupling coefficient, referred to as the axion-photon coupling. 
Despite its weak coupling, the interaction offers a direct way to search for axion in laboratory experiments, simultaneously placing a tight constraint on  $\gag$~\cite{Irastorza_Redondo2018,Sikivie_2021review,Adams:2022pbo}. Also, astronomical observations provide clues about axion by modulating observed photons, altering the lifetimes of astronomical objects, and triggering new phenomena~\cite{DiLuzio:2020wdo,Galanti:2022ijh}. The absence of these signatures, therefore, allows us to constrain the coupling parameter $\gag$.

In this {\it Letter}, we consider a novel search for axion DM,  utilizing the terrestrial EM fields in the extremely low-frequency (ELF) band. Coupled with the static geomagnetic field, coherently oscillating axions  permeating the entire Earth can generate monochromatic EM waves at a frequency corresponding to the axion mass. In particular, EM waves generated near the Earth's surface at frequencies of $1\,{\rm Hz}\lesssim f\lesssim 100$\,Hz are confined between the Earth's surface and the lower ionosphere, making the natural environment of the Earth in the ELF range a powerful window for exploring ultralight axions (see Ref.~\cite{Beadle_etal2024} for a possible window at the MHz band). The use of such an environment has been pointed out in~\cite{Arza:2021ekq}. Focusing especially on the frequency of $f\lesssim10^{-2}$\,Hz, they have placed a constraint on $\gag$ in the mass range, $2\times10^{-18}\,{\rm eV}\lesssim \mass\lesssim 7\times10^{-17}\,{\rm eV}$, utilizing the magnetometer network on Earth.  
Recently, the constraint has been extended to heavier masses, $\mass\lesssim 4\times10^{-15}\,{\rm eV}$, with high-sampling measurement data~\cite{Friel:2024shg}. There are other efforts constraining the coupling with direct measurements in the mass range, $10^{-15}\,{\rm eV}$--$10^{-12}\,{\rm eV}$~\cite{CAST:2017uph,CAST:2024eil,Sulai:2023zqw,Oshima:2023csb,Bloch:2023wfz}. 

Here, our prime focus is the EM fields at the frequencies of $f=0.3-30$\,Hz, whose wavelength is comparable or longer than the Earth circumference. In contrast to the frequency range of $f\lesssim1$\,Hz studied in Refs.~\cite{Arza:2021ekq,Friel:2024shg}, resonant behavior of EM waves is expected, and reliable predictions requires a proper theoretical modeling that accounts for atmospheric conductivity. The present study allows us to explore an axion signature or to put a tight bound on the axion-photon coupling across a mass range of $10^{-15}\,{\rm eV}-10^{-13}\,{\rm eV}$.

%%%%%%%%%%%%%%%%%%%%%%%%%%%%%%%%%%%%%%%%%%%%%%%%%%%%%%%%
\noindent{\bf Axion-induced signal in terrestrial $B$-fields.} In the presence of axions, the Maxwell equations are modified, and there appear effective charge and current through the interaction $\mathcal{L}_{\rm int}$. If axions comprise the DM, they have non-relativistic velocity, $v_{\rm DM}\sim10^{-3}$, and the terms involving spatial gradient, $|\nabla a|\sim\mass v_{\rm DM}a$, are generally suppressed, compared to those involving the time derivative $|\partial_t a|\sim \mass a$, where we adopt the natural units, and set the speed of light to unity.  Then, the relevant modification appears only through the effective current, given by $\bfJ_{\rm eff}=-\gag\,(\partial_t a)\,\bfB$. Further, the axions are predominantly coherent, and exhibit an oscillatory behavior such that $a\propto e^{-i\,\mass t}$. This implies that when coupled with a static magnetic field, the axion fields produce an alternating current with a frequency of $f_{\rm a}=\mass/(2\pi)\simeq2.4\,(\mass/10^{-14}\,{\rm eV})$\,Hz, which can serve as a source of monochromatic EM waves at the same frequency $f_{\rm a}$ via the modified Amp\'ere-Maxwell law. Assuming the characteristic scale of $R$, its magnetic field amplitude $\bfB_{\rm a}$ is estimated roughly by equating the terms in the Amp\'ere-Maxwell law, $\nabla\times\bfB\sim\bfJ_{\rm eff}$. This yields $|\bfB_{\rm a}|\sim \gag \mass a_0\,R\,|\bfB_0|$, with $\bfB_0$ and $a_0$ being respectively the amplitudes of static magnetic and axion fields. 

In the terrestrial environment, the geomagnetic field serves as a representative static and global magnetic field, characterized by a dipole configuration, with its strength typically of $|\bfB_{\rm geo}|\sim 25-65\,\mu$T \cite{IGRF13_2021}. Setting $\bfB_0$ to $\bfB_{\rm geo}$ and assuming that the size of magnetic field is comparable to the Earth radius, $R_{\rm E}=6371$\,km, we have
%%%%%%%%%%%%%%%%%%
\begin{align}
    |\bfB_{\rm a}|& \sim 0.3\,{\rm pT} \left(\frac{\gag}{10^{-10}\,{\rm GeV}^{-1}}\right)
    \left(\frac{\rho_{\rm DM}}{0.3\,{\rm GeV}\,{\rm cm}^{-3}}\right)^{1/2}\left(\frac{R}{R_{\rm E}}\right)\left(\frac{|\bfB_{\rm geo}|}{50\,\mu{\rm T}}\right).
    \label{eq:rough_estimation_axion_B_field}
\end{align}
%%%%%%%%%%%%%%%%%%
Here, $\rho_{\rm DM}$ is the local DM, which is related to the quantity $(\mass a)^2/2$ averaged over the timescales longer than the coherence time $T_{\rm coh}\sim 2\pi/(\mass v_{\rm DM}^2)$. Since the bandwidth of the produced EM waves is estimated to be $\Delta f/f_{\rm a}\sim v_{\rm DM}^2\sim10^{-6}$, the induced EM fields appear as a sharp frequency spike, persisting over the coherence time.

Although the amplitude at Eq.~\eqref{eq:rough_estimation_axion_B_field} is more than seven orders of magnitude smaller than that of the geomagnetic fields, it is still detectable with high-precision commercial magnetometers. Further, at $f=0.3-30$\,Hz, the major background noise is a random superposition of transient EM waves (e.g., \cite{Barr_etal2000,Simoes_etal2012}), which can be discriminated from the static axion signal having a sharp spectral line by using a long-term monitoring data of terrestrial magnetic fields.

Let us compute the induced EM fields in more realistic setup by properly accounting for the geometric configuration of the Earth's magnetic field and the atmospheric conductivity. The latter is crucial to obtain a finite EM amplitude propagating near the Earth's surface especially at the frequencies of our interest, where the resonant behavior is expected. We are interested in searching for the axion-induced signal through the measurement of magnetic fields at the Earth's surface. Considering its non-uniform nature, the induced magnetic fields are generally expanded with the vector spherical harmonics, $\bfPhi_{\ell m}\equiv \bfr\times\nabla Y_{\ell m}$, where $\bfr$ is the radial vector from the Earth's center and $Y_{\ell m}$ is the scalar spherical harmonics.
The signal satisfying the divergence-free condition is expressed as \cite{TNH_Theory,NTH_DataAnalysis} 
%%%%%%%%%%%%%%%%%%%%%%%%%%%%%%%%%%%%%%%%%%%%%%%%%%%%%%%%
\begin{align}
    \bfB_{\rm a}(r,\hat{\Omega};\mass) &= g_{\rm a\gamma}  a\sum_{\ell,m}b_{\ell}(\mass\,r)\,C_{\ell m}\,\bfPhi_{\ell m}(\hat{\Omega}),
    \label{eq:axion-induced_B-field}
\end{align}
%%%%%%%%%%%%%%%%%%%%%%%%%%%%%%%%%%%%%%%%%%%%%%%%%%%%%%%%
where the $\hat{\Omega}=(\theta,\phi)$ stands for the geographic coordinate on the Earth surface, and the $C_{\ell m}$ is the harmonic coefficient of the magnetic scalar potential $V$, which describes the geomagnetic field through $\bfB_{\rm geo}=-\nabla V$. The function $b_\ell$ is the radial mode function that characterizes the response of the induced magnetic fields to the axions at the radius $r$. At the Earth's surface $r=R_{\rm E}$, its amplitude depends solely on the axion mass, $\mass$. Introducing the dimensionless variable $x \equiv\mass r$, this is obtained by solving the following equation with appropriate boundary conditions, and taking only the homogeneous part of the solutions \cite{TNH_Theory}:

%%%%%%%%%%%%%%%%%%%%%%%%%%%%%%%%%%%%%%%%%%%%%%%%%%%%%%%%%%%%%%%%%%%%%%%
\begin{figure}[t]

\vspace*{-0.5cm}

\begin{center}
\includegraphics[width=10cm,angle=0]{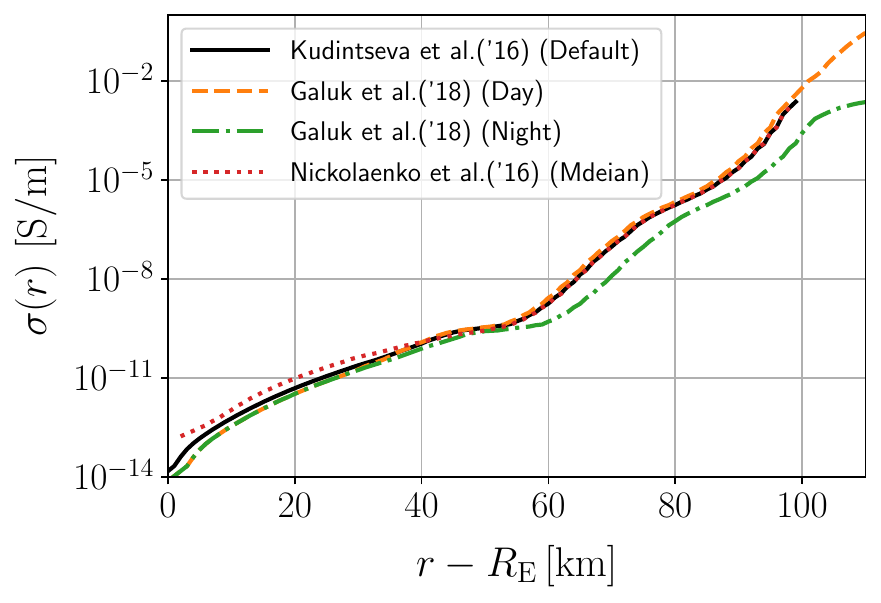}
    \end{center}

\vspace*{-0.5cm}

\caption{Atmospheric conductivity profiles as a function of altitude above the Earth's surface, $r-R_{\rm E}$. 
While the solid curve is taken from Ref.~\cite{Kudintseva_etal2016},  dashed, dash-dotted, and dotted curves correspond to the day-side and night-side profiles of Ref.~\cite{Galuk_etal2018} and the median profile of Ref.~\cite{Nickolaenko_etal2016}, respectively. 
Note that these profiles are provided as tabulated data with different altitude ranges.
\label{fig:conductivity_profiles}
}
\end{figure}
%%%%%%%%%%%%%%%%%%%%%%%%%%%%%%%%%%%%%%%%%%%%%%%%%%%%%%%%%%%%%%%%%%%%%%%
Introducing the dimensionless variable $x \equiv \mass r$, this is obtained by solving the following equation under the boundary condition that the tangential electric field vanishes at the Earth's surface and the homogeneous solution approaches an upward-propagating wave near the lower ionosphere ($r\sim R_{\rm E}+100\,\unit{km}$), and retaining only the homogeneous part of the solution (see Ref.~\cite{TNH_Theory} for more details): 

%%%%%%%%%%%%%%%%%%%%%%%%%%%%%%%%%%%%%%%%%%%%%%%%%%%%%%%%
\begin{align}
    &\Bigl\{\frac{d}{dx}\Bigl(\frac{1}{n^2}\frac{d}{dx}\Bigr)+1-\frac{\ell(\ell+1)}{n^2x^2}\Bigr\}(x\,b_{\ell})
%    \nonumber
%    \\
%    &
    =-\frac{(\mass R_{\rm E})^{\ell+2}}{\ell\,x^{\ell}}.
    \label{eq:mode_eq}
\end{align}
%%%%%%%%%%%%%%%%%%%%%%%%%%%%%%%%%%%%%%%%%%%%%%%%%%%%%%%%
Note that Eq.~\eqref{eq:mode_eq} is derived by using the wave equation for the electric field alongside the Maxwell-Faraday equation. 
Here, the influence of the atmospheric conductivity is incorporated into the complex refractive index $n$ through $n^2=1+i\,(\sigma/\mass)$, where the conductivity $\sigma$ increases monotonically with the radial coordinate $r$, from $10^{-14}$ to $10^{-3}$\,S/m between the Earth's surface and the lower ionosphere, as shown in Fig.~\ref{fig:conductivity_profiles} (see Ref.~\cite{TNH_Theory} for details).
%%%%%%%%%%%%%%%%%%%%%%%%%%%%%%%%%%%%%%%%%%%%%%%%%%%%%%%%%%%%%%%%%%%%%%%
\begin{figure}[t]

\vspace*{-0.5cm}

\begin{center}
\includegraphics[width=10cm,angle=0]{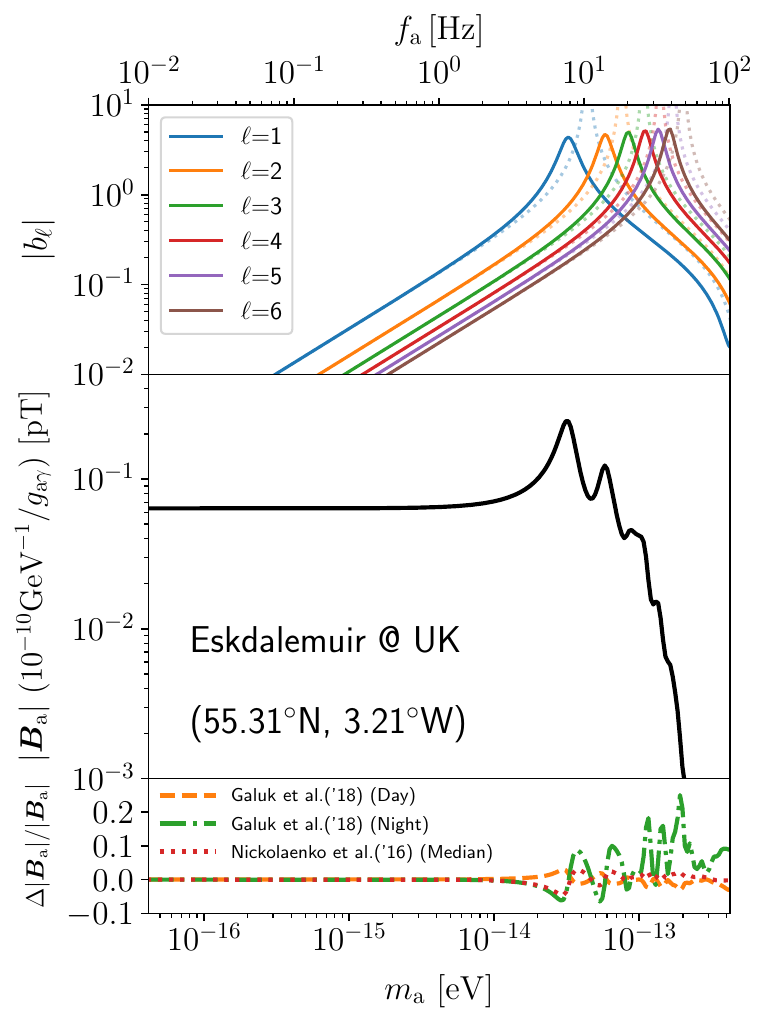}
    \end{center}

\vspace*{-0.5cm}

\caption{{\it Top:} Radial mode function $b_{\ell}$ for the lowest six multipoles, evaluated at the Earth's surface $r=R_{\rm E}$. The solid curves are the results adopting the conductivity profile of Ref.~\cite{Kudintseva_etal2016}, which  are independent of geographic location, and plotted against the axion mass $\mass$
(lower) and the induced EM wave frequency $f_{\rm a}$ (upper). For comparison, the dotted curves indicate the mode functions in an idealized setup that neglects the atmospheric conductivity. 
{\it Middle:} Expected amplitude of the axion-induced magnetic field at Eskdalemuir observatory, normalized by the axion-photon coupling strength of $10^{-10}\,$GeV$^{-1}$. Following Eq.~\eqref{eq:axion-induced_B-field}, this is obtained by summing the mode functions from the top panel, weighted by the harmonic coefficients of geomagnetic field and vector spherical harmonics, and evaluating at the observatory’s location.
{\it Bottom:} Impact of different conductivity profiles on the axion-induced magnetic field, shown as fractional differences. Dashed, dash-dotted, and dotted curves correspond to those shown in Fig.~\ref{fig:conductivity_profiles}, relative to the reference model in the middle panel.
\label{fig:Response_axion_B-field_Eskdalemuir}
}
\end{figure}
%%%%%%%%%%%%%%%%%%%%%%%%%%%%%%%%%%%%%%%%%%%%%%%%%%%%%%%%%%%%%%%%%%%%%%%

Adopting the conductivity profile in Ref.~\cite{Kudintseva_etal2016} (black solid curve in Fig.~\ref{fig:conductivity_profiles}), we solve the radial mode equation by treating the radial dependence of $\sigma$ or $n$ as a sum of spatially constant segments. This approach allows us to analytically solve for $b_\ell$, while satisfying the boundary conditions at the Earth's surface and the lower ionosphere \cite{TNH_Theory}. 
The top panel of Figure~\ref{fig:Response_axion_B-field_Eskdalemuir} plots the resultant mode functions for the lowest six multipoles at the Earth's surface, given as a function of $\mass$ or the frequency $f_a$ of the induced EM signal. There is a prominent peak structure in each multipole, known as the Schumann resonance~\cite{Schumann1952a, Schumann1952b} (see also  Refs.~\cite{Jackson_1998,Nickolaenko_Hayakawa2002,Simoes_etal2012,Nickolaenko_Hayakawa2013}), which is caused by EM waves trapped between the Earth's surface and the lower ionosphere, forming a resonant cavity. The peak frequencies are close to $\sqrt{\ell(\ell+1)}/(2\pi\,R_{\rm E})\simeq 7.46\sqrt{\ell(\ell+1)}$\,Hz, as predicted in the idealized case of a perfectly conducting Earth-ionosphere cavity (dotted curves), although the actual peak positions depicted as solid curves are systematically lower than these values by 20-30\% due to losses in the propagation of EM waves. As a result, the resonant peaks have a finite amplitude, approximately $b_{\ell} \sim 5$
for $\ell=1-6$, unlike the idealized case where the amplitudes exhibit a divergent behavior. 

%%%%%%%%%%%%%%%%%%%%%%%%%%%%%%%%%%%%%%%%%%%%%%%%%%%%%%%%
%%%%%%%%%%%%%%%%%%%%%%%%%%%%%%%%%%%%%%%%%%%%%%%%%%%%%%%%
\noindent
{\bf Predictions of induced magnetic fields.}  
%%%%%%%%%%%%%%%%%%%%%%%%%%%%%%%%%%%%%%%%%%%%%%%%%%%%%%%%
%%%%%%%%%%%%%%%%%%%%%%%%%%%%%%%%%%%%%%%%%%%%%%%%%%%%%%%%
Given the mode function $b_\ell$ above, the prediction of induced magnetic field strength is obtained from Eq.~\eqref{eq:axion-induced_B-field} by convolving further the information on the geomagnetic field configuration ($C_{\ell m}$) and a specific geographic location on the Earth. The coefficients $C_{\ell m}$ are described by the International Geomagnetic Reference Field (IGRF) model with the harmonic coefficients up to $\ell=13$ \cite{IGRF13_2021}. 

The middle panel of Fig.~\ref{fig:Response_axion_B-field_Eskdalemuir} shows the predicted magnetic field amplitude at the Eskdalemuir observatory, UK, $(55.31^\circ\,{\rm N},\,3.21^\circ\,{\rm W})$, where long-term monitoring data up to $100\,$Hz is available (see below). Here, we assume that the axions constitute the DM having the local density of $\rho_{\rm DM}=0.3\,{\rm GeV}\,{\rm cm}^{-3}$, 
with the coupling parameter of $g_{a\gamma}=10^{-10}$\,GeV$^{-1}$. 
The induced magnetic field is predicted to exhibit a peak at $\mass\sim3\times10^{-14}$\,eV, corresponding to the peak frequency $f_{\rm a}\sim7.8$\,Hz for the radial mode function of $\ell=1$. Additionally, a low-frequency plateau of $|\bfB_{\rm a}|\sim\mathcal{O}(10^{-1})$\,pT is observed, consistent with Ref.~\cite{Arza:2021ekq}. On the other hand, at $\mass\gtrsim10^{-13}$\,eV, its amplitude is sharply suppressed, reflecting the dipole nature of the geomagnetic fields, where the higher multipoles are sufficiently small. 

The spectral features seen in the middle  panel of Fig.~\ref{fig:Response_axion_B-field_Eskdalemuir} appear typical regardless of the geographic location, though the overall amplitude varies mostly with latitude, increasing near the equator, which reflects a toroidal field configuration \cite{TNH_Theory}. While these predictions rely on the specific atmospheric conductivity model of Ref.\cite{Kudintseva_etal2016}, we find that the choice of conductivity profile only has a mild effect. As shown in the bottom panel of Fig.\ref{fig:Response_axion_B-field_Eskdalemuir}, even though conductivity models of Refs.~\cite{Nickolaenko_etal2016,Galuk_etal2018} differ by nearly an order of magnitude around the lower ionosphere (see Fig.~\ref{fig:conductivity_profiles}), the actual impact on the magnetic field at the Earth's surface remains small, reaching at most $\sim20$\% in a regime where the magnetic field is already suppressed.

%%%%%%%%%%%%%%%%%%%%%%%%%%%%%%%%%%%%%%%%%%%%%%%%%%%%%%%%
%%%%%%%%%%%%%%%%%%%%%%%%%%%%%%%%%%%%%%%%%%%%%%%%%%%%%%%%
\noindent{\bf Terrestrial magnetic field data.}
%%%%%%%%%%%%%%%%%%%%%%%%%%%%%%%%%%%%%%%%%%%%%%%%%%%%%%%%
%%%%%%%%%%%%%%%%%%%%%%%%%%%%%%%%%%%%%%%%%%%%%%%%%%%%%%%%
Based on the predictions above as a theoretical template, we use the publicly-available magnetic field data to search for the axion signature. Specifically, 
the data used here are those measured at the Eskdalemuir Observatory~\cite{Beggan-Musur-2018} during  September 1, 2012 to November 4, 2022, which are maintained at the British Geological Survey~\cite{BGS-data}. The data consist of two channels for the North-South and East-West directions (CH1 and CH2) with the sampling rate of $100\unit{Hz}$, both of which are calibrated within the range of $0.001$--$100\unit{Hz}$. While the orientation of the measured magnetic fields helps distinguish the signal from noise, the systematics and stability are not well-understood. Therefore, we use only the power spectral density (PSD), computed from the two channels, CH1 and CH2. 

The basic data analysis is summarized as follows. First, a continuous dataset spanning more than one month is extracted from the full data, which is then divided into segments of length $T_{\rm seg} = 8\,{\rm hrs}$. Second, each data segment is Fourier-transformed, and its PSD is computed. The frequency bin size of each segment is $\Delta f_{\rm a}\simeq 1/T_{\rm seg} \simeq 3.5\times10^{-5}$\,Hz, roughly corresponds to the bandwidth of the axion-induced signal at $f_{\rm a}=35$\,Hz, above which the signal decays rapidly as shown in the middle panel of Fig.~\ref{fig:Response_axion_B-field_Eskdalemuir}. Axion-induced signals at $f_{\rm a}\lesssim35\,\unit{Hz}$ are well within a single frequency bin. Considering the coherence time, the segment size of 8 hrs may not be an optimal choice but is chosen so as to avoid unknown-cause short-duration transient noise with huge amplitude of $\sim 10^8\unit{pT}$ that happens irregularly. Due to the transient noise, the fraction of the data segments with good quality is roughly $\sim 10$\% of the total number of the data segments, $N_{\rm seg}=9909$.

\begin{figure}[t]
\begin{center}
%\hspace*{-0.3cm}
\includegraphics[width=12cm]{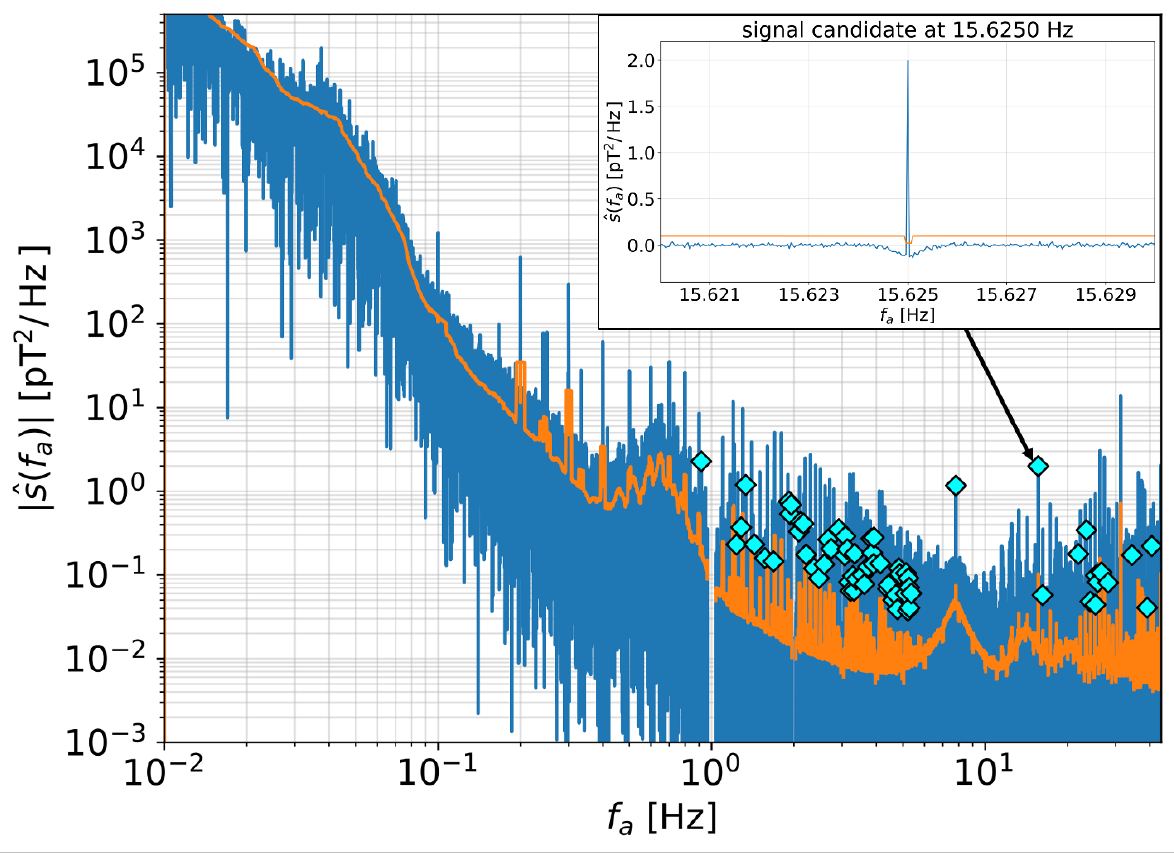}
\end{center}
\caption{Weighted-average differential PSD of the magnetic fields, stacked over all eight-hour data segments with noise-based weights and subtracted the smooth spectrum, $\hat{s}(f_{\rm a})$ in blue. The standard deviation is depicted in orange. The axion signal candidates (SNR$>3$) are marked by diamonds in cyan. The inset highlights the signal candidate with the highest SNR $(13.342$).
\label{fig:differential-averaged-PSD-8h}}
\end{figure}

%%%%%%%%%%%%%%%%%%%%%%%%%%%%%%%%%%%%%%%%%%%%%%%%%%%%%%%%
%%%%%%%%%%%%%%%%%%%%%%%%%%%%%%%%%%%%%%%%%%%%%%%%%%%%%%%%
\noindent{\bf Constraining axion-photon coupling.} 
%%%%%%%%%%%%%%%%%%%%%%%%%%%%%%%%%%%%%%%%%%%%%%%%%%%%%%%%
%%%%%%%%%%%%%%%%%%%%%%%%%%%%%%%%%%%%%%%%%%%%%%%%%%%%%%%%
Given the PSD of the magnetic fields, we stack them with the weight of the inverse of noise variance computed from the spectrum at surrounding frequencies. Then we subtract the smooth component, obtained by applying a quadratic Butterworth filter with a sampling frequency of $0.05 \unit{Hz}$. This allows the axion-induced line signals to add constructively, while reducing other components (magnetometer noise variance and the Schumann resonance spectrum). Dividing by the sum of the inverse-of-noise-variance weight for data segments, we obtain the weighted average differential PSD, $\hat{s}(f_{\rm a})$. The procedure above still leads to numerous lines at integer-multiple frequencies, the exact origin of which is unknown. Since these are likely artificial, we remove the data surrounding these frequencies ($\pm1000$ bins). The resultant PSD is shown in Fig.~\ref{fig:differential-averaged-PSD-8h}.
%%%%%%%%%%%%%%%%%%%%%%%%%%%%%%%%%%%%%%%%%%%%%%%%%%%%%%%%
%%%%%%%%%%%%%%%%%%%%%%%%%%%%%%%%%%%%%%%%%%%%%%%%%%%%%%%%

The search for the axion-induced signal consists of evaluating the signal-to-noise ratio (SNR) for each frequency bin and selecting the candidates that exceed a certain threshold. The standard deviation of the weighted-average differential PSD is calculated from the surrounding $\pm200$ bins, excluding the three bins around $f=f_{\rm a}$, as shown by the orange curve in Fig.~\ref{fig:differential-averaged-PSD-8h}.
We regard this as noise amplitude to estimate the SNR for the axion signal at $f_{\rm a}$, and evaluate the SNR from $0.01$\,Hz, below which the number of samples is insufficient, to $44$\,Hz, above which the estimated spectrum becomes less reliable because it is close to the Nyquist frequency. 

We then identify signal candidates by requiring (i) ${\rm SNR} > 2$ in all-year data, (ii) ${\rm SNR}$ in each year's data larger than the SNR threshold (2 in our case) weighted by the sum of the inverse of noise variances in the year, ensuring persistence over the entire observation period, and (iii) that their frequencies are at least $10^{-3}\unit{Hz}$ away from multiples of $0.05\unit{Hz}$. The third condition is for excluding likely artificial line noises. We found $342$ candidates (see also \cite{NTH_DataAnalysis} for signal candidates with different SNR thresholds). These candidates appear in a single frequency bin, consistent with the sharpness of an axion signal. There are 1 candidate with significant SNR (13.342) and 31 candidates with SNR$>5$ for the all-year data. In Fig.~\ref{fig:differential-averaged-PSD-8h}, representative candidates of SNR$>3$ are depicted by cyan-filled diamonds. Lacking enough information to veto them, we retain them as potential axion signal candidates. 

We note that below $1$ Hz, Ref. ~\cite{Friel:2024shg} reported three signal candidates with modest significance using SuperMAG 1-second sampling data. Although our analysis identified five candidates with SNR$ > 2$, none of them matched those reported in Ref.~\cite{Friel:2024shg}. Given that our measurement sensitivity is comparable to theirs\footnote{To be precise, our sensitivity is slightly worse than theirs for the candidate at $0.2630$Hz identified by Ref.~~\cite{Friel:2024shg}.}, the signal candidates identified below $1$ Hz in both analyzes can be ruled out.

Apart from the signal candidates mentioned above, no axion-like signal was found. We can therefore place upper limits on the axion coupling strength in the frequency range of $0.01-43.97$\,Hz. The 95\% C.~L. upper limit on the axion-photon coupling strength is obtained from
%%%%%%%%%%%%%%%%%%%%%%%%%%%%%%%%%%%%%%%%%%%%%%%%%%%%%%%%%%%%%
\begin{align}
    \int_{-\infty}^{\hat{s}_{\rm obs}(m_{\rm a})} {\mathrm d} \hat{s}\, p[\hat{s}(m_{\rm a})|g_{\rm a\gamma}] = 0.05 \;.
\end{align}
%%%%%%%%%%%%%%%%%%%%%%%%%%%%%%%%%%%%%%%%%%%%%%%%%%%%%%%%%%%%%
The $\hat{s}_{\rm obs}$ is the observed value of the weighted-average differential PSD, which can be converted to the axion coupling strength by equating it with $\langle\hat{s}_{\rm a}(m_{\rm a})\rangle \equiv 2 T_{\rm seg} |\bfB_{\rm a} (m_{\rm a})|^2$ (see Ref.~\cite{NTH_DataAnalysis} for the derivation) and using the theoretical curve in the middle panel of Fig.~\ref{fig:Response_axion_B-field_Eskdalemuir}. The $p[\hat{s}(m_{\rm a})|g_{\rm a\gamma}]$ is the probability density distribution of the weighted-average differential PSD in the presence of axions, constructed from data $\hat{s}$ of the surrounding $\pm200$ bins (excluding the three bins around $f=f_{\rm a}$) and shifted by the axion signal $\hat{s}_{\rm a}$ at $f=f_{\rm a}$. 

\begin{figure}[t]
\begin{center}
\includegraphics[width=12cm,angle=0]{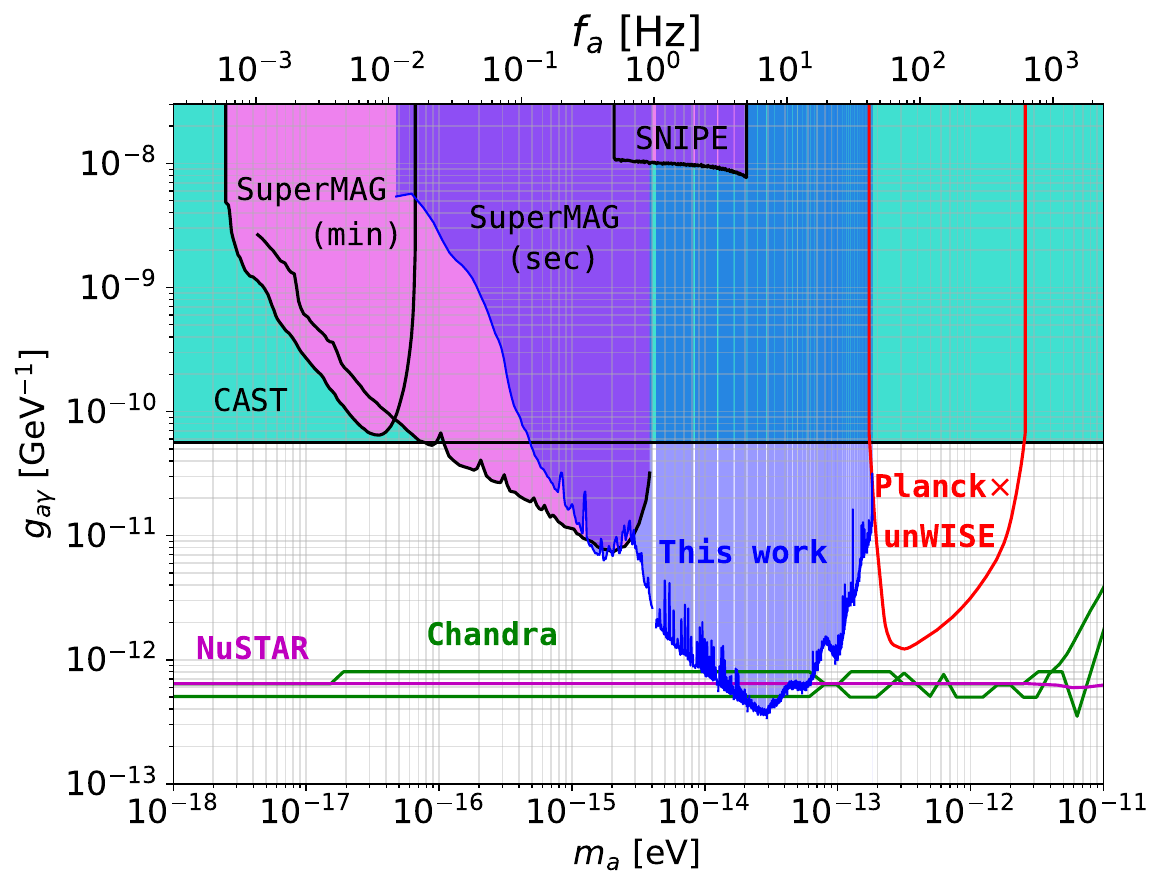}
\end{center}
%\hspace*{0.5cm}
\caption{Constraint (95\% C.~L.) on the axion-photon coupling $\gag$ from the long-term monitoring data of magnetic fields at Eskdalemuir observatory (this work, blue) except for the notched frequency bands. Other excluded regions are from SuperMAG~\cite{Arza:2021ekq,Friel:2024shg}, SNIPE~\cite{Sulai:2023zqw}, Planck and unWISE blue galaxy sample~\cite{Goldstein:2024mfp}, CAST~\cite{CAST:2017uph,CAST:2024eil}, 
Chandra~\cite{Reynes:2021bpe,Reynolds:2019uqt}, and NuSTAR~\cite{Ning:2024eky}. 
The data are taken from \cite{AxionLimits}. The constraints by direct means are filled and those by astrophysical model-dependent means are not filled.}
\label{fig:constraint_axion_photon_coupling}
\end{figure}

Figure~\ref{fig:constraint_axion_photon_coupling} shows the upper limit on the axion coupling we obtained (in blue), along with other observation bounds. Tightest limit is obtained around an axion mass of $3 \times 10^{-14}\unit{eV}$, which is improved by about two orders of magnitude compared to the CAST result ~\cite{CAST:2017uph,CAST:2024eil}, and exceeds the limit from 
astrophysical X-ray observations by Chandra~\cite{Reynes:2021bpe,Reynolds:2019uqt,Dessert:2021bkv,Ning:2024ozs} and NuSTAR~\cite{Ning:2024eky}. 
In the lower mass range, our constraint is weakened because of fixing the data segment size to 8 hrs and using broader frequency resolution, irrespective of the coherence time of axions. In the higher mass range above $10^{-14}\unit{eV}$, the constraint is also weakened due to the suppression of the magnetic field's response to axions.

%%%%%%%%%%%%%%%%%%%%%%%%%%%%%%%%%%%%%%%%%%%%%%%%%%%%%%%%%%%%%%%%%%%%%%%
%%%%%%%%%%%%%%%%%%%%%%%%%%%%%%%%%%%%%%%%%%%%%%%%%%%%%%%%%%%%%%%%%%%%%%%

%%%%%%%%%%%%%%%%%%%%%%%%%%%%%%%%%%%%%%%%%%%%%%%%%%%%%%%%
%%%%%%%%%%%%%%%%%%%%%%%%%%%%%%%%%%%%%%%%%%%%%%%%%%%%%%%%
\noindent{\bf Conclusions and outlook.} 
%%%%%%%%%%%%%%%%%%%%%%%%%%%%%%%%%%%%%%%%%%%%%%%%%%%%%%%%
%%%%%%%%%%%%%%%%%%%%%%%%%%%%%%%%%%%%%%%%%%%%%%%%%%%%%%%%
In this {\it Letter}, we properly accounted for atmospheric conductivity to predict terrestrial EM waves induced by coherently oscillating axions, enabling a search for axion DM signals in public magnetic field data. We set the most stringent direct upper bound on the axion-photon coupling parameter at $10^{-15}\,{\rm eV}\lesssim \mass\lesssim 10^{-13}\,{\rm eV}$, and identify potential signal candidates with significant SNR for further study. Follow-up ground-based experiments such as DANCE \cite{Obata:2018vvr} and those employing twisted anyon cavities~\cite{Bourhill:2022alm}, as well as X-ray observations by Athena \cite{Sisk-Reynes:2022sqd} will be crucial to confirm or rule out these candidates. While we focus on axions as a representative ultralight DM, our methodology applies to other DM coupled to EM fields, such as dark photon DM \cite{Nelson_Scholtz2011,Fabbrichesi_etal2020}, whose kinetic mixing can be constrained using the same dataset~\cite{Fedderke_etal2021,Fedderke_etal2021_superMAG,NNTH_Darkphoton}.

Finally, we address the so-called stochastic effect~\cite{Centers_etal2019,Lisanti_etal2021}, which arises from the superposition of field modes with different frequencies and random phases, potentially weakening the upper limit on the axion-photon coupling. Ref.~\cite{Nakatsuka:2022gaf} shows that such an effect has a negligible impact on the upper bound if the the measurement time is more than ten times the coherence time. In our analysis, this condition is met for axion masses greater than $2\times 10^{-15}\unit{eV}$ (see~\cite{NTH_DataAnalysis} for a detailed discussion). Therefore, the assumption of a constant axion amplitude is justified, and our derived constraint remains robust above this threshold.

\bigskip 
\noindent{\bf Acknowledgements} 
We are grateful to C. Beggan for  kindly providing the data at the Eskdalemuir observatory and for his valuable comments. 
This work was supported in part by JSPS KAKENHI Grant Numbers JP20H05861, JP23K20844, and JP23K25868 (AT), JP23K03408, JP23H00110, and JP23H04893 (AN), JP21K03580 (YH). Numerical computation was carried out partly at the Yukawa Institute Computer Facility.

\bibliographystyle{apsrev4-1}
%\bibliography{ref}
\input{references.bbl}

\end{document}